\documentclass[preprint,preprintnumbers,amsmath,amssymb]{revtex4}

\usepackage{graphicx}           % Include figure files
\usepackage{dcolumn}            % Align table columns on decimal point
\usepackage{bm}                 % bold math
\usepackage{hyperref}

\begin{document}

%\preprint{APS/123-QED}

\title{Non-resonant dot-cavity coupling and its applications in resonant quantum dot spectroscopy}

\author{S. Ates$^{(1)\star}$, S. M. Ulrich$^{(1)\star}$, A. Ulhaq$^{(1)}$, S. Reitzenstein$^{(2)}$, A. L\"offler$^{(2)}$,
S. H\"ofling$^{(2)}$, A. Forchel$^{(2)}$, \& P. Michler$^{(1)}$}

\affiliation{$^{(1)}$ Institut f\"ur Halbleiteroptik und
Funktionelle Grenzfl\"achen, Universit\"at Stuttgart, Allmandring
3, 70569 Stuttgart, Germany,}

\affiliation{$^{(2)}$ Technische Physik, Universit\"at W\"urzburg,
Am Hubland, 97074 W\"urzburg, Germany.\\ \\
$^{\star}$~Email correspondence: s.ates@ihfg.uni-stuttgart.de;
s.ulrich@ihfg.uni-stuttgart.de }

\maketitle

%\pacs{78.67.Hc, 71.70.Ej, 78.55.Cr, 71.35.-y}

\textbf{Promising solid-state single-photon sources and cavity
quantum electrodynamic schemes have been realized on the basis of
coupled quantum dot and micro-/nanocavity systems
\cite{Michler.Kiraz:2000,Moreau.Robert:2001}. Recent experimental
studies on the single quantum dot (QD) level showed a pronounced
emission at the cavity resonance even for strongly detuned
dot-cavity systems
\cite{Hennessy.Badolato:2007,Press.Forchel:2007,Badolato.Imamoglu:2008}.
This behaviour is indicative of a complex light-matter interaction
in a semiconductor well beyond the widely used two-level
emitter-cavity schemes. Different mechanisms such as
photon-induced 'shake-up' processes in charged quantum dots
\cite{Kaniber.Finley:2008}, dephasing processes
\cite{Naesby.Moerk:2008,Yamaguchi.Asano.Noda:2008,Auffeves.Gerard.arXiv:2009}
and phonon-mediated processes \cite{Tarel.Savona.arXiv:2008} are
currently discussed to understand the experimentally observed
features. A well prepared and clearly defined experimental
situation is therefore mandatory to gain a thorough understanding
of the responsible physical mechanisms behind the non-resonant
dot-cavity coupling.}

\textbf{Here we present experimental investigations on the
non-resonant dot-cavity coupling of a single quantum dot inside a
micro-pillar where the dot has been resonantly excited in the
s-shell, thereby avoiding the generation of additional charges in
the QD and its surrounding. As a direct proof of the pure single
dot-cavity system, strong photon anti-bunching is consistently
observed in the autocorrelation functions of the QD and the mode
emission, as well as in the cross-correlation function between the
dot and mode signals. Strong Stokes and anti-Stokes-like emission
is observed for energetic QD-mode detunings of up to $\sim 100$
times the QD linewidth. Furthermore, we demonstrate that
non-resonant dot-cavity coupling can be utilized to directly
monitor and study relevant QD s-shell properties like
fine-structure splittings, emission saturation and power
broadening, as well as photon statistics with negligible
background contributions.}

\textbf{Our results open a new perspective on the understanding
and implementation of dot-cavity systems for single-photon
sources, single and multiple quantum dot lasers, semiconductor
cavity quantum electrodynamics, and their implementation, e.g. in
quantum information technology \cite{Bouwmeester.Ekert:2000}.}

%%%%% Introduction %%%%%

High $Q$-factor nano- and micro-cavities can enhance or suppress
the spontaneous emission of photons, e.g. from a quantum dot,
coupled to a well defined mode by the Purcell effect
\cite{Vahala:2003}. For very high-$Q$ (i.e. weakly damped)
cavities the spontaneous emission can even become a reversible
process so that quantum entanglement of radiation and matter
becomes possible, in the so-called \textit{strong coupling} regime
\cite{Vahala:2003}. Resonantly coupled single quantum dot nano-
and micro-cavity systems, i.e. with QD and cavity mode in
resonance, have been realized both in the \textit{weak coupling}
\cite{Michler.Kiraz:2000,Santori.Fattal:2002} and the
\textit{strong coupling} regime
\cite{Reithmaier.Sek:2004,Yoshie.Scherer:2004,Peter.Senellart:2005}.
Recent experimental results also show significant emission at the
cavity resonance even if the single quantum dot is not in
resonance with the cavity mode
\cite{Hennessy.Badolato:2007,Press.Forchel:2007,Badolato.Imamoglu:2008}.
Similar observations have also been reported from nano-cavity
laser structures with only few QDs as the active medium
\cite{Strauf.Hennessy.ea:2006}. This so-called
\textit{non-resonantly coupled emission} mechanism is not well
understood and controversially discussed in the literature:
Kaniber et al. \cite{Kaniber.Finley:2008} suggested that the
experimentally observed coupling between a single QD and a
photonic crystal cavity mode is mediated by photon-induced
'\textit{shake-up}'-like processes in charged quantum dots. In
their theoretical work
\cite{Naesby.Moerk:2008,Auffeves.Gerard.arXiv:2009}, Naesby et al.
and Auff\`{e}ves et al. demonstrated that \textit{dephasing}
shifts the emission intensity towards the cavity frequency,
whereas Tarel and Savona \cite{Tarel.Savona.arXiv:2008} showed the
important role of the \textit{electron-acoustic-phonon
interaction} for understanding the emission properties. From the
experimental point of view it is desirable to study the coupling
mechanism via purely resonant excitation of the QD s-shell in
order to obtain a better understanding of the underlying physics.
However, in all \textit{experimental} studies such a
\textit{purely resonant excitation} of the QD s-shell has not been
performed so far. The non-resonantly coupled system we use is
based on individual self-assembled (In,Ga)As/GaAs QDs embedded in
a high-quality micro-pillar cavity. To ensure single exciton
generation and to suppress background emission we use
\textit{purely resonant s-shell excitation} of an uncharged QD at
low temperatures ($T < 30$~K). Signatures of strong emission
coupling are observed for both negative and positive spectral
detunings between a single QD and the fundamental cavity mode.
Furthermore, we demonstrate that the cavity mode emission can be
conveniently used to monitor essential QD s-shell properties in
high detail while avoiding the complications involved with direct
investigations of the resonance fluorescence and/or transmission
(e.g. stray light) and reflection experiments (e.g.
nano-apertures, noisy background, or demand on high setup
sensitivity).

%%%%%%%%%%%%%%%%%%%%%%%%%%% Figure 1 %%%%%%%%%%%%%%%%%%%%%%%%%%%%%%%%%

For all investigations in the current work, a special orthogonal
geometry of sample excitation and emission detection was used
(\textbf{Fig.~1b}; for details, see methods section) in order to
optically address and study the emission characteristics of single
QDs in selected micro-pillar structures. \textbf{Figure~1a} shows
a characteristic low-temperature ($T = 22$~K) photoluminescence
spectrum of a micro-pillar cavity under p-shell excitation of a
single QD. Two emission lines are visible and are identified as
excitonic emission from the QD ($\sim 1.3574$~eV) and fundamental
mode emission ($\sim 1.3577$~eV) of the pillar cavity. In order to
study the physical origin of the mode emission in more detail,
photon auto- and cross-correlation measurements between the two
lines have been performed and are displayed in \textbf{Figs.~1c -
e}. All correlation measurements display pronounced photon
anti-bunching. The \textit{auto-correlation} of the single QD
(trace~\textbf{c}) displays perfect anti-bunching (i.e.
$g^{(2)}(\tau = 0) = 0.0(2)$; deconvoluted value) as expected for
a single quantum emitter \cite{Michler.Imamoglu:2000}.
Surprisingly, also the mode itself exhibits a strongly reduced
$g^{(2)}(0)$ value of $0.20\,\pm\,0.02$, demonstrating that the
observed mode emission
is mainly ($\sim\,90\,\%$) originating from a single quantum emitter. %(Dies ist z.B. bei Finley et al nicht der Fall)%
The signature of strong photon anti-bunching even in the
\textit{cross-correlation} of these two lines (with $g^{(2)}(0) =
0.10\,\pm\,0.02$) demonstrates a pronounced \textit{non-resonant
dot-cavity coupling} between the same single QD and the
$\sim\,280\,\mu$eV detuned pillar mode as was also consistently
observed by other groups
\cite{Hennessy.Badolato:2007,Press.Forchel:2007} under
non-resonant barrier pumping and p-shell excitation, respectively.
The effect is not yet clarified from a theoretical perspective,
thus making further detailed studies indispensable to gain a
deeper understanding of the underlying processes. Furthermore, as
we show in the following, this non-resonant dot-cavity coupling
can be utilized to effectively transfer and 'monitor' the
information of various relevant QD s-shell characteristics onto a
nearly background-free coupled mode channel.

%%%%%%%%%%%%%%%%%%%%%%%%%%% Figure 2 %%%%%%%%%%%%%%%%%%%%%%%%%%%%%%%%%

In \textbf{Figs.~2a} and \textbf{b}, the effect of non-resonant
QD-cavity coupling is demonstrated for both \textit{positive} and
\textit{negative} QD-mode detunings $\Delta E = +200\,\mu$eV
(-380~$\mu$eV) at $T = 10$~K (29~K). Excitation laser frequency
scans are performed on the s-shell resonance of a single quantum
dot, and the resulting spectra are displayed as a function of the
laser-exciton detuning $\delta$. Under consecutive variation of
the laser frequency, two prominent effects can be clearly
observed. First, a strong increase in the composite signal of the
QD s-shell emission and scattered laser light demonstrates the
excitation of \textit{resonance fluorescence} from the quantum
emitter. Furthermore, in parallel with the generation of resonance
fluorescence, significant emission signal is also observed from
the detuned pillar mode. Worth to note, this effect is found over
a wide temperature range, with positive and negative QD-mode
detunings of up to $\sim 100\,\times$ the single quantum dot
s-shell emission FWHM (or up to $\sim 4\,\times$ the fundamental
mode linewidth). In \textbf{Fig.~2c} we plot the \textit{relative}
mode emission intensity ratio, normalized to the sum of QD and
mode emission, as a function of QD-mode detuning $\Delta E$. Each
data point represents the conditions of fully resonant excitation
($\delta = 0$) into the QD s-shell at a constant power level ($P_0
= 300$~nW). Interestingly, a pronounced increase of the normalized
mode intensity is traced with increasing temperature -- in spite
of the increasing QD-mode detuning $\Delta E(T)$. A large relative
coupling efficiency of $\sim 72\%$ is observed from the collected
signal. These results strongly support the recent suggestion that
\textit{phonon-mediated processes} may play a dominant role for
the non-resonant dot cavity coupling
\cite{Naesby.Moerk:2008,Tarel.Savona.arXiv:2008}. In particular,
due to the resonant excitation process of single neutral excitons
in the QD we can exclude photon-induced 'shake-up' processes
\cite{Kaniber.Finley:2008} in our samples.

%%%%%%%%%%%%%%%%%%%%%%%%%%% Figure 3 %%%%%%%%%%%%%%%%%%%%%%%%%%%%%%%%%

\textbf{Figures~3a - d} display results of laser resonance scan
series (\textbf{a, b}: $T = 18$~K, \textbf{c, d}: $T = 26$~K) on
the s-shell of the single QD in pillar~2, in direct comparison
with the emission characteristics of the non-resonantly coupled
pillar mode. In the corresponding top and bottom traces, each data
point represents the integrated intensity derived from Lorentzian
line fits to the mode (plots \textbf{a, c}) and the QD s-shell
emission (plots \textbf{b, d}), respectively. As a consequence of
resonant excitation the direct QD s-shell signal is composed of
resonance fluorescence and laser stray light, reflected as a small
non-vanishing background level for large detunings $\delta$ from
the resonance (traces \textbf{b, d}). The data sets in \textbf{a}
and \textbf{b} observed at $T = 18$~K reveal distinct doublet
structures with similar linewidths and energy splittings ($\Delta
E_{FSS} = 11\,\pm\,0.3\,\mu$eV). The doublet structure of the
excitonic line is caused by the electron-hole exchange interaction
in an asymmetric QD \cite{Bayer.Ortner:2002}. In \textbf{Figs.~3c}
and \textbf{d} the known effect of power broadening
\cite{Stufler.Zrenner:2004} of the excitonic line under resonant
excitation is compared with the coupled mode emission. A
significant line broadening is clearly visible under stepwise
increasing laser powers, revealing high quantitative accordance
between the FWHM of the $\mu$-PL lines in the data sets \textbf{c}
and \textbf{d}. Please note that these power-dependent
measurements have been performed at $T = 26$~K, and as a
consequence the fine structure splitting remains unresolved due to
the significant thermal broadening of the line. Our results
demonstrate that relevant QD s-shell properties like the excitonic
\textit{fine structure}, the \textit{relative emission strengths}
and \textit{spectral widths}, as well as the \textit{power
broadening} of these $\mu$-PL lines can be accurately monitored
via the almost background-free mode emission.

%%%%%%%%%%%%%%%%%%%%%%%%%%% Figure 4 %%%%%%%%%%%%%%%%%%%%%%%%%%%%%%%%%

Moreover, in the following we demonstrate that the QD
\textit{emission saturation} and the \textit{photon statistics}
under resonant s-shell excitation can also be conveniently
measured via the coupled micro-pillar mode. \textbf{Figure~4a}
depicts results of a power series on a resonantly (s-shell) pumped
single QD for an emitter-mode detuning of $\Delta E =
+200\,\mu$eV, monitored via the fundamental cavity mode. We
observe the characteristic saturation behavior of a resonantly
pumped single quantum dot \cite{Stufler.Zrenner:2004}. The
experimental data has been fitted by a theoretical model
\cite{Flagg.Muller:Shih:2009} to describe the emission saturation
of a resonantly excited two-level system (see also methods
section). Under the given experimental conditions we obtain fit
values of $T_1 = 670\,\pm\,50$~ps for the radiative lifetime and
$T_2 = 460\,\pm\,50$~ps for the emission coherence. With respect
to the radiative lifetime of the decay, high conformity is found
from a comparison with the results of independent time-resolved
spectroscopy measurements (TCSPC; see inset in \textbf{Fig.~4a}),
yielding a value of $T_1 = 650\,\pm\,20$~ps.

\textbf{Figure~4b} displays the auto-correlation measurement of
another resonantly (s-shell) pumped single QD (pillar~3) with a
large emitter-mode detuning of $\Delta E = +440\,\mu$eV. Under
purely resonant ($\delta = 0$) pumping of the QD s-shell, the
correlation was taken on the coupled fundamental cavity mode
signal (see inset spectra). For the regime of low excitation
powers applied in the experiment, a single correlation-'dip' but
no oscillatory behaviour is expected for the $g^{(2)}(\tau \sim
0)$ trace around zero delay (see methods section). Indeed, very
pronounced photon anti-bunching is found in the measured data from
the pillar mode with $g^{(2)}(0) \approx 0.26$, derived from a
convoluted fit to the experimental trace (bold red line) under
consideration of the temporal resolution $t_{IRF}$ of our setup.
Applying the same fit parameters (i.e. $T_1$ and $T_2$ under the
given conditions) as before, but assuming now \textit{full}
temporal resolution (as shown by the de-convoluted dashed curve),
even the ideal case of $g^{(2)}(0) = 0.0(2)$ for a single quantum
emitter is achieved. In clear contrast to the
\textit{background-limited} conditions of \textit{p-shell
excitation} discussed in \textbf{Fig.~1e}, the data of
\textbf{Fig.~4b} nicely demonstrates the absence of background
light, i.e. pure single-photon emission under the given conditions
of strictly resonant excitation into the QD s-shell, monitored via
a non-resonantly coupled pillar mode.

As a concluding remark, we envision that the demonstrated
technique of resonance fluorescence emission 'monitoring' via
efficient non-resonant emitter-mode coupling in a micro-resonator
represents a very versatile and powerful tool for fundamental
studies on the single quantum-dot level. The large coupling
efficiencies demonstrate the exciting potential of this technique
for resonantly pumped single-photon sources with negligible
background. Under proper conditions, even a high degree of photon
indistinguishability can be anticipated
\cite{Auffeves.Gerard.arXiv:2009}. In general, the technique
should be applicable to a large variety of state-of-the-art
micro-cavity geometries, thus allowing for cavity-QED studies with
unprecedented detail.

\subsection*{Methods Summary}

We use single (In,Ga)As/GaAs quantum dots (QDs) in all-epitaxial
high-quality vertical cavity micro-pillar cavities to study in
detail the effect of non-resonant emitter-mode coupling at
cryogenic temperatures $T < 30$~K. For these investigations a
novel sample design and measurement geometry is chosen, providing
optical access to individual micro-cavity pillar structures close
to a cleaved edge of the sample structure. Optical laser
excitation in the lateral plane of QD growth is combined with
emission detection along the vertical pillar axis. This technique
provides an enhanced separation of resonance fluorescence from
scattered laser signal with sufficient signal-to-noise contrast.
Resonant s-shell excitation of single QDs is provided by a
continuous-wave (cw) narrow-band (500~kHz) tunable Ti:Sapphire
laser with 30~GHz scan range. By controlled utilization of the
different temperature-dependence of cavity mode spectra and QD
s-shell states, a systematic study of the non-resonant
emitter-mode coupling effect in 'Stokes' and 'Anti-Stokes'
configurations is enabled. The high efficiency of this coupling
mechanism is used for a background-free 'monitoring' of numerous
essential QD s-shell properties via the coupled (and detuned)
pillar mode emission.

\subsection*{Acknowledgements}

The authors gratefully acknowledge financial support by the DFG
research groups FOR~730 ''Positioning of Single Nanostructures --
Single Quantum Devices'' and FOR~485 ''Quantum Optics in
Semiconductor Nanostructures''. We also thank M.~Emmerling and
A.~Wolf for expert sample preparation.

%%%%%%%%%% References %%%%%%%%%%

\bibliographystyle{naturemag}
\bibliography{Bib_Ates}

\newpage

%%%%%%%% Figure Captions %%%%%%%

\textbf{Figure~1: Photon statistics of a non-resonantly coupled
QD-micro-pillar cavity system.} \textbf{a,} Low-temperature ($T =
22$~K) micro-photoluminescence spectra of the s-shell of a single
(In,Ga)As/GaAs QD coupled to the spectrally detuned fundamental
mode of a micro-pillar cavity (pillar~1). In this case, optical
excitation of the QD is applied via laser absorption in the first
excited state (p-shell), $\sim 22$~meV higher in energy (not
shown). Please note that in these $\mu$-PL spectra the resolution
is limited to $\sim 35\,\mu$eV. \textbf{b,} Orthogonal geometry of
excitation and detection in our $\mu$-PL experiments:
Micro-pillars close to a cleaved sample edge are individually
addressed by a focused laser within the lateral QD growth plane;
$\mu$-PL emission is detected along the vertical micro-pillar
axis. \textbf{c -- e,} Second-order photon auto-correlation
measurements on the pure QD or mode emission (traces \textbf{c}
and \textbf{e}, respectively), together with the cross-correlated
QD-mode signal (\textbf{d}) observed from pillar~1. The bold
(dashed) lines represent theoretical fits to the data (open
circles), convoluted (de-convoluted) with respect to the detector
response time $t_{IRF} = 400$~ps, respectively. \\ \\

\textbf{Figure~2: Laser $\delta$-frequency scanning over a single
quantum dot s-shell resonance for different dot-cavity detunings
$\Delta E$.} \textbf{a} and \textbf{b,} Selected spectra from
detailed cw-laser resonance scan series of a coupled QD-cavity
system (pillar~2) for the case of \textit{positive} ($\Delta E =
+200 \,\mu$eV at $T = 10$~K) and \textit{negative} ($\Delta E =
-380 \,\mu$eV at $T = 29$~K) spectral QD-mode detuning. In full
consistence with the experimental findings from several studied
coupled QD-cavity systems, either case of QD-mode detuning $\Delta
E$ indicates a strong non-resonant coupling mechanism between the
resonantly excited QD fluorescence and the micro-pillar mode
emission. As demonstrated in series \textbf{a} and \textbf{b},
maximum QD resonance fluorescence under s-shell excitation (i.e.
laser detuning $\delta = 0$~GHz) consistently appears with maximum
\textit{'Stokes'} ($\Delta E > 0$) or \textit{'anti-Stokes'}
($\Delta E < 0$) mode signal. \textbf{c,} Normalized mode emission
ratio as a function of QD-mode detuning $\Delta E(T)$,
demonstrating increasing mode emission with increasing detuning
(i.e. temperature $T$) from the coupled single QD.

\newpage

\textbf{Figure~3: Monitoring single QD s-shell emission properties
via a non-resonantly coupled pillar mode.} \textbf{a, b,}
High-resolution s-shell resonance scan (frequency step width:
$\sim 270$~MHz) on the single QD in micro-pillar~2, taken at $T =
18$~K under low excitation power ($P_0 = 100$~nW). Laser tuning
over the s-shell clearly reveals a structural asymmetry-induced
\textit{fine structure} doublet signature with $\Delta E_{FSS} =
11.3 \pm 0.3\,\mu$eV in the spectrally integrated emission
intensity (trace~\textbf{b}), which is composed of QD resonance
fluorescence and scattered laser signal. From a comparison with
the mode signal (detuning $\Delta E = +180\,\mu$eV) in trace
\textbf{a}, detailed 'mapping' of all spectral features of the
non-resonantly coupled QD emission is observed. \textbf{c, d,}
Excitation power-dependent s-shell resonance scan series of the
same QD (pillar~2), demonstrating the effect of
\textit{power-broadening} for a resonantly pumped two-level system
(trace \textbf{d}). A high level of consistence with corresponding
data from the coupled mode emission in \textbf{c} is found. Note
that due to an increased sample temperature of $T = 26$~K,
emission fine structures remain unresolved in this case. \\ \\

\textbf{Figure~4: QD emission saturation monitoring via a coupled
micro-pillar mode.} \textbf{a,} Power series of the resonantly
(s-shell) pumped single QD inside pillar~2 for a QD-mode detuning
of $+ 200\,\mu$eV. Data points represent the spectrally integrated
emission signal of the non-resonantly coupled micro-pillar mode.
The denoted values of $T_1$ (radiative lifetime) and $T_2$
(coherence length) are temperature-sensitive fit parameters to the
data. The value of $T_1$ has been independently verified by
time-resolved spectroscopy (TCSPC; see inset figure) under p-shell
excitation. \textbf{Monitoring of single-photon generation from a
resonantly excited QD. b,} Photon auto-correlation measurement on
the fundamental mode emission of another coupled QD-mode system
(pillar~3; $\Delta E = +440\,\mu$eV) under weak ($P_0 = 300$~nW)
and purely resonant s-shell excitation of the dot. Despite the
large detuning from the emitter, the mode signal fully reflects
the \textit{ideal} case of background-free single-photon emission
from the coupled QD. The bold (dashed) lines represent theoretical
fits to the $g^{(2)}(\tau)$ data, convoluted (de-convoluted) with
respect to the temporal setup resolution.

\newpage

%%%%%%%%%%%% Figures %%%%%%%%%%%%

\textbf{Figure~1}

\begin{figure}[!ht]
\begin{centering}
\includegraphics[width=8cm]{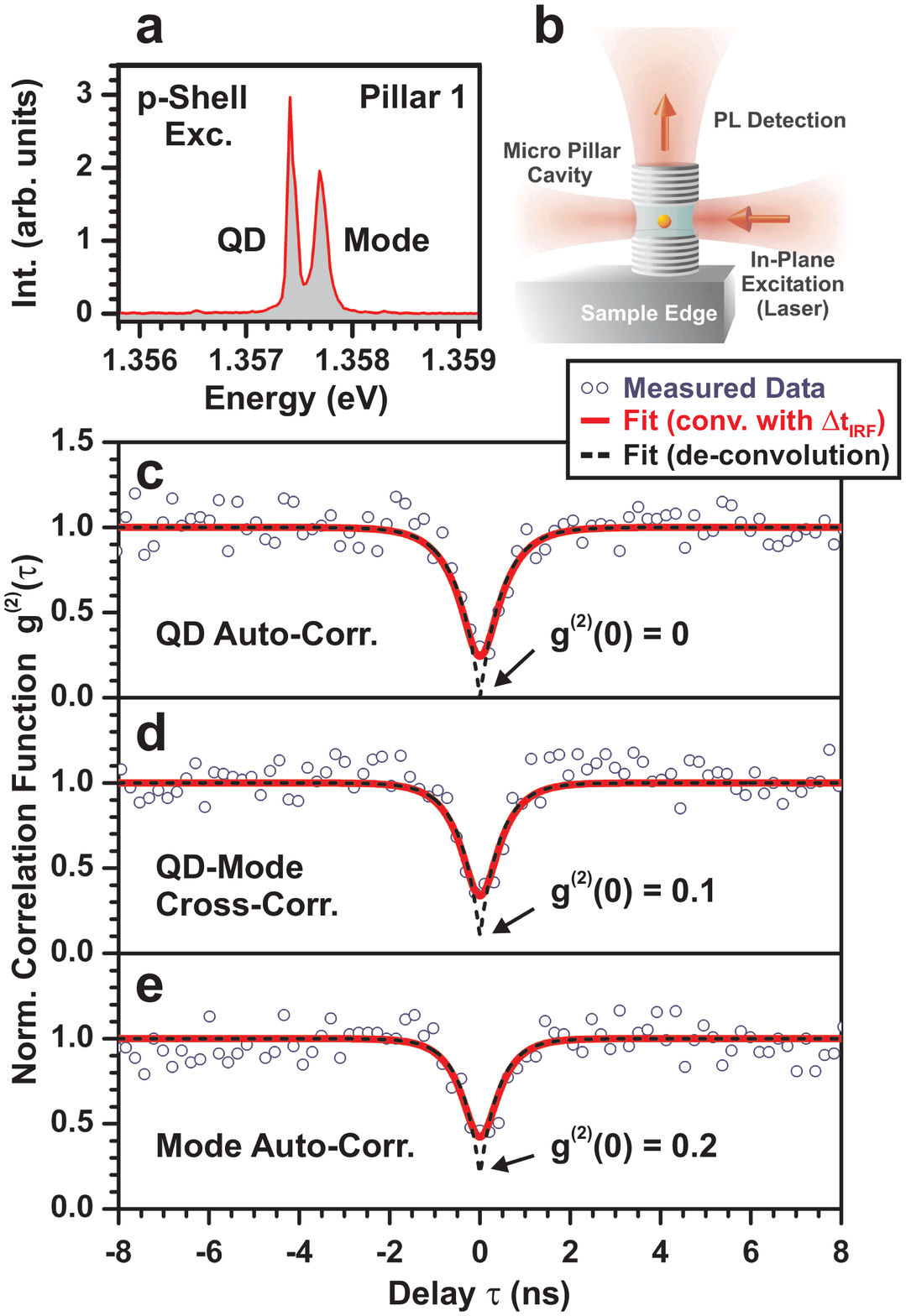}
\label{fig:1}
\end{centering}
\end{figure}

\vspace{0.5cm}
\textbf{Figure~2}

\begin{figure}[!ht]
\begin{centering}
\includegraphics[width=12.5cm]{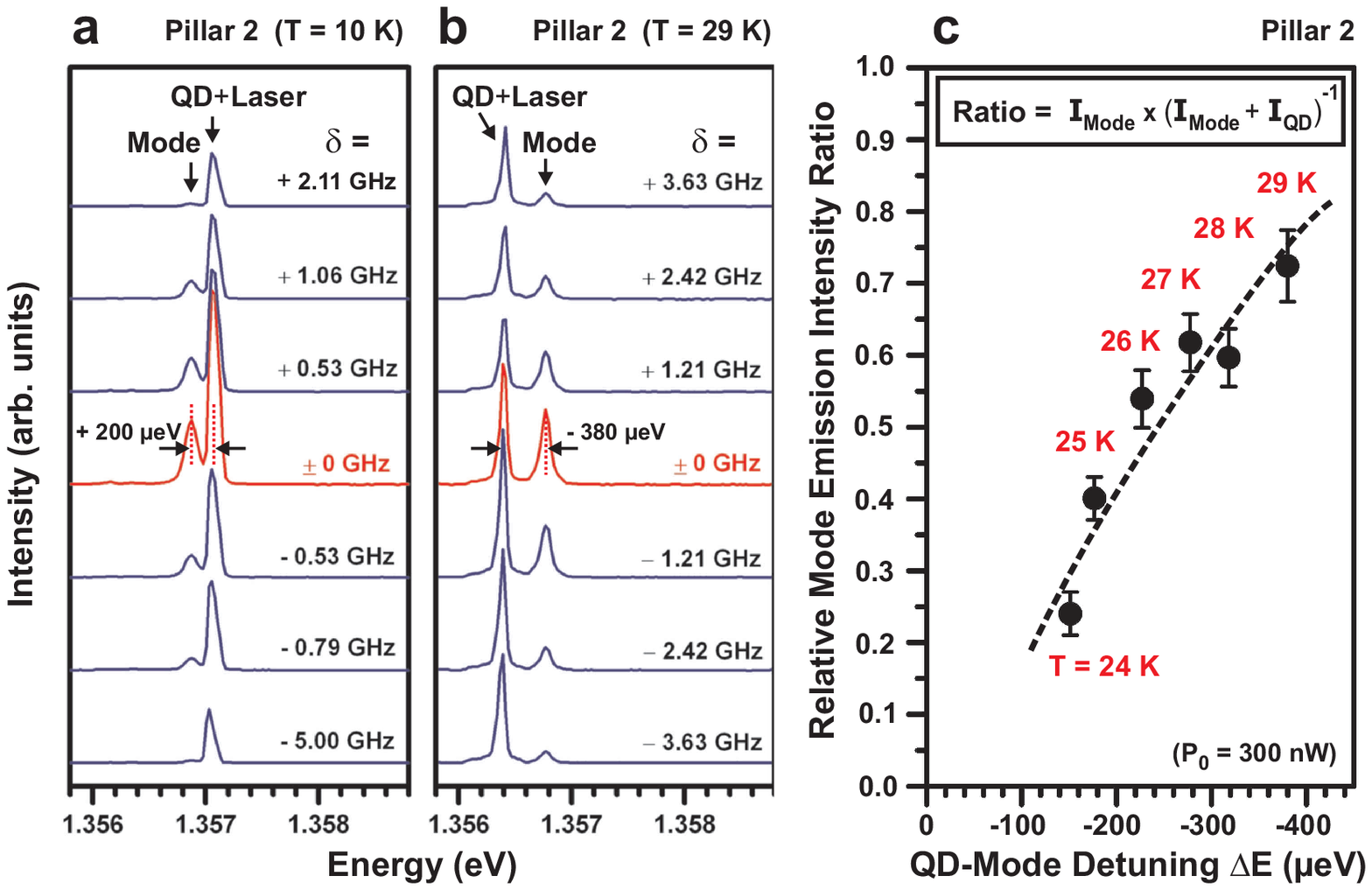}
\label{fig:2}
\end{centering}
\end{figure}

\newpage
\textbf{Figure~3}

\begin{figure}[!ht]
\begin{centering}
\includegraphics[width=11.5cm]{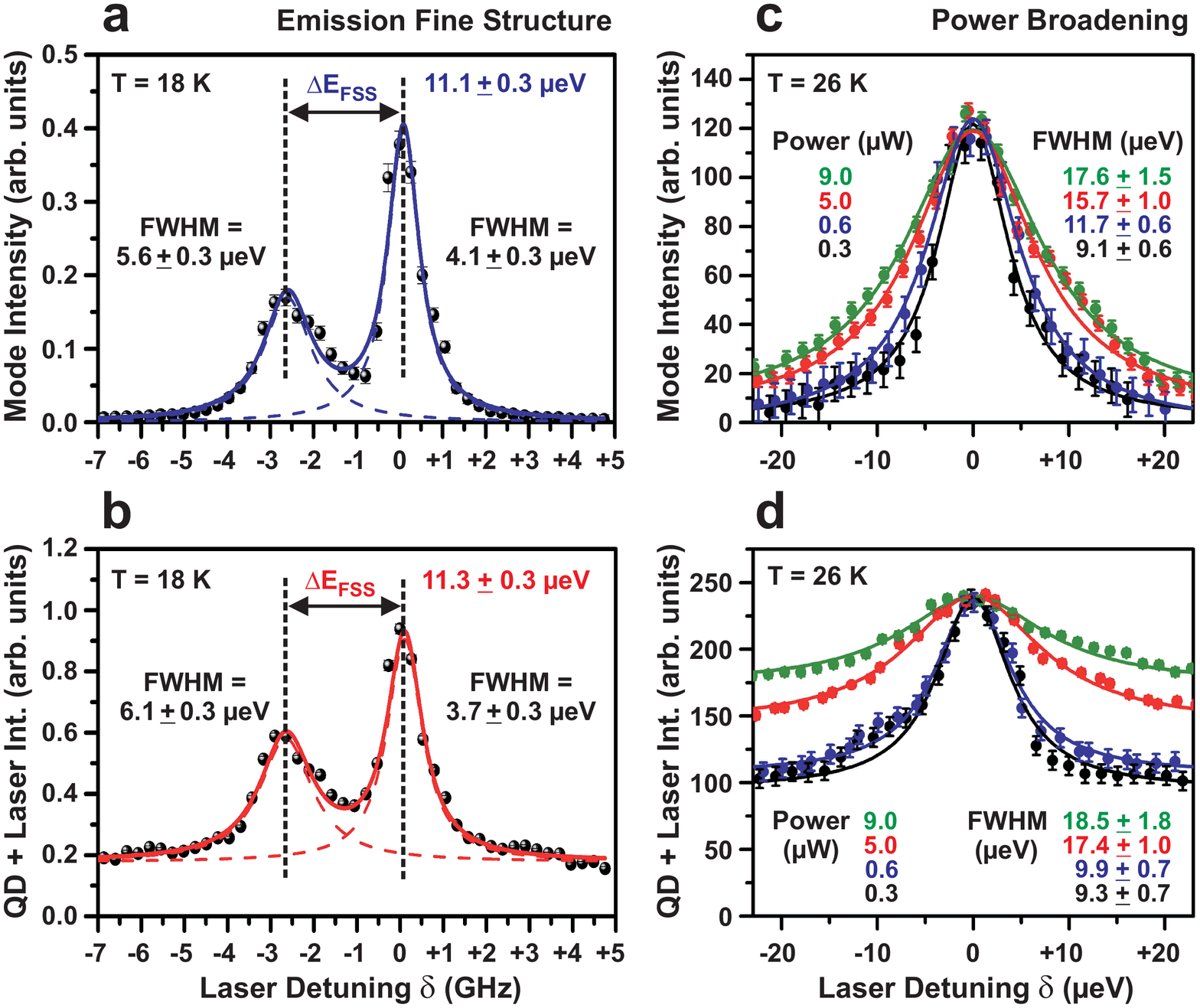}
\label{fig:3}
\end{centering}
\end{figure}

\vspace{0.5cm}
\textbf{Figure~4}

\begin{figure}[!ht]
\begin{centering}
\includegraphics[width=12.5cm]{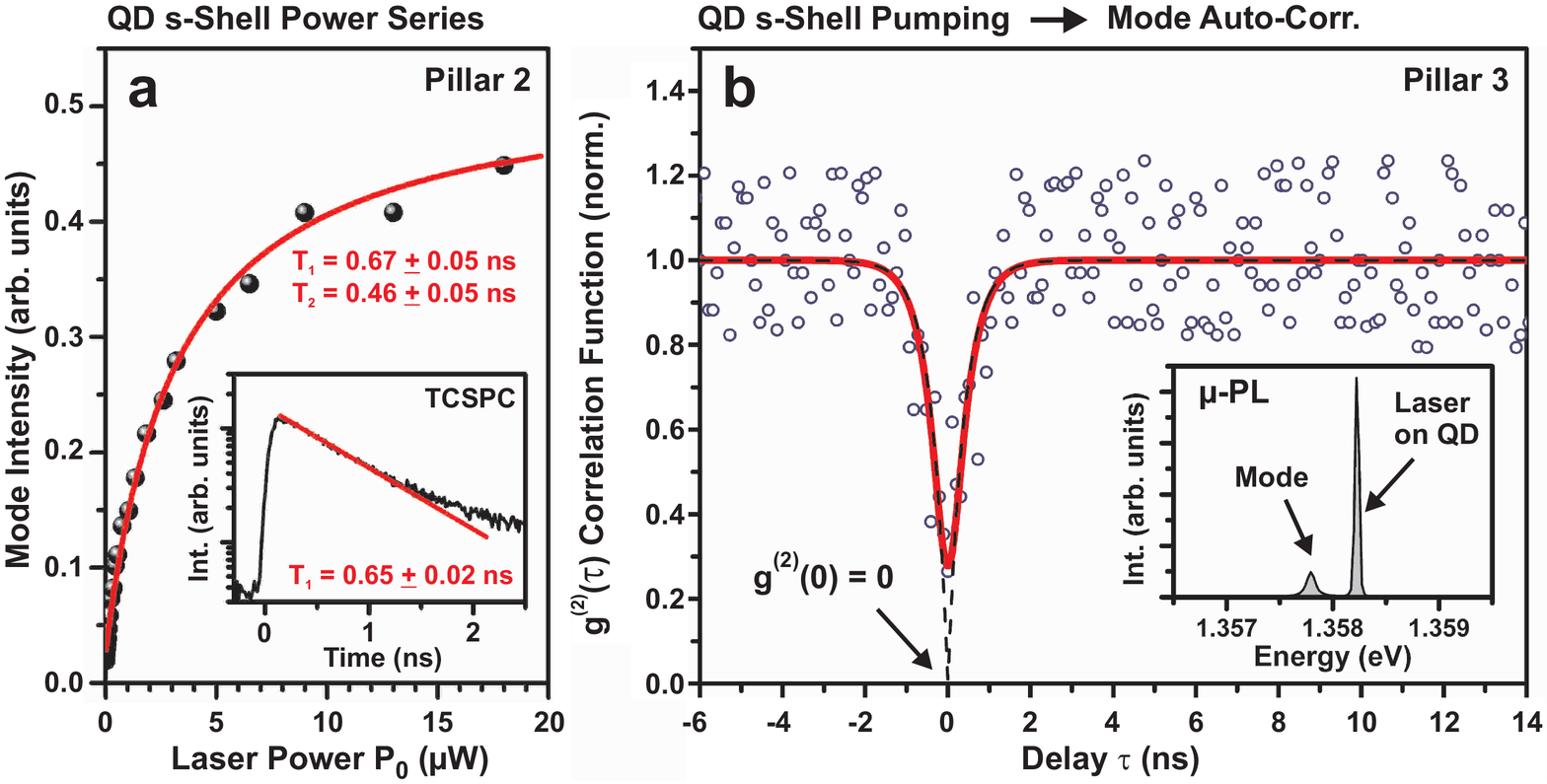}
\label{fig:4}
\end{centering}
\end{figure}

\newpage

\section*{Methods}

\subsection*{Sample Growth and Preparation}

Our samples were grown by molecular beam epitaxy on a GaAs
substrate. The initial planar sample structure consists of a
single layer of self-assembled (In,Ga)As QDs (indium content $\sim
45\%$; lateral QD density $\sim 10^{10}$~cm$^{-2}$) which is
positioned at the center of a $1\,\lambda$-thick GaAs barrier
layer. On top and below the central cavity layer, 26 and 30
periods of alternating $\lambda\,/\,4$-thick layers of AlAs/GaAs
were grown to form distributed Bragg reflectors (DBRs),
respectively. Micro-pillar resonator arrays were fabricated by a
combination of electron-beam lithography and reactive ion etching,
details of which are described in \cite{LoefflerForchel:2005}. For
our studies we have cleaved the structured sample along the major
orientation axes of the micro-pillar arrays, leaving a lateral
distance of less than $5\,\mu$m between the outer row of pillars
and the sample edge for optimum optical access. A single pillar
structure contains only 150-250 QDs on average which are
additionally spectrally spread due to their inhomogeneous
distribution within the whole dot ensemble (peak position at 1.37
eV; FWHM $\sim$~100~meV).

\subsection*{Experimental Setup}

In our studies, the sample was mounted in a cold-finger He-flow
cryostat and stabilized to temperatures of $T \leq 30$~K with
$\pm\,0.5$~K accuracy. Optical excitation of individual
micro-pillars was provided by either a tunable narrow-band
($\Delta \nu$~=~500~kHz (FWHM)) continuous-wave (cw) Ti:Sapphire
ring laser or by a mode-locked tunable Ti:Sapphire pulse laser
(76~MHz repetition rate; $\sim 2$~ps pulse width) used in
time-resolved micro-photoluminescence (TCSPC). A
computer-controlled 3D-precision scanning stage equipped with a
50x microscope (fiber-coupled, SLWD, NA = 0.45) was horizontally
adapted to the cryostat unit and used for selective optical
excitation of QDs within their lateral growth plane.
\textbf{Fig.~1b} schematically depicts the orthogonal geometry of
excitation and emission detection (50x microscope, SLWD, NA =
0.45) along the vertical micro-pillar axis. This setup allows for
a repetitive and long-term stable addressing of individual
micro-pillars under strong suppression of scattered laser stray
light. Stray light suppression was additionally improved by use of
adjustable pin-holes (PH) within the detection path towards the
monochromator (not shown), which limit the area of detection in
the focal plane to an effective spot diameter of $\sim\,2\,\mu$m.
The collected light was spectrally dispersed by one or two
1200~line/mm grating spectrometers each equipped with a $\ell
N$-cooled high-sensitivity charge-coupled device (CCD) camera for
time-integrated photon acquisition. To investigate the photon
emission statistics, a Hanbury Brown and Twiss-type (HBT) setup
\cite{HanburyBrownTwiss-B:1956} for $g^{(2)}(\tau)$ second-order
auto-correlation measurements was used. In these correlation
measurements, the collected and spectrally pre-filtered photon
stream is divided by a 50/50 non-polarizing beam splitter into two
symmetric paths, each equipped with an avalanche photo diode (APD)
for efficient single-photon detection. Using a time-to-amplitude
converter to transform the measured time separations $\tau =
t_{stop} - t_{start}$ of photon coincidences between the
designated 'start' and 'stop' APD channels, the full correlation
is stored into a multichannel analyzer. Limited by the response
time of the APDs, our HBT setup provides a full temporal
resolution of $\Delta t_{res} \approx \,400$~ps. As discussed in
the main text, a correct interpretation of the quality of
single-photon generation in terms of 'anti-bunching' in
$g^{(2)}(\tau)$ measurements demands data de-convolution with the
instrumental temporal response function (here: Gaussian profile of
FWHM $\Delta t_{IRF}$).

\subsection*{Theoretical Data Analysis}

%Emission Saturation:

According to the theoretical expectations for a \textit{resonantly
pumped two-level system} under zero laser detuning $\delta = 0$
\cite{Muller.Flagg:2007,Flagg.Muller:Shih:2009}, the emission rate
(i.e. intensity) of spontaneous recombination from the excited
state obeys the relation
\begin{equation}\label{eqn:resInt}
I_{res}(P_0) \propto \frac{1}{2} \frac{\Omega^2 \cdot T_1/T_2}
{T_2^{-2} + \Omega^2 \cdot T_1/T_2} \quad .
\end{equation}
According to this expression, emission saturation is expected for
a regime of strong excitation where $\Omega^2 \gg 1/\sqrt{\left (
T_1 \, T_2 \right )}$. In order to enable an explicit fit to our
experimental example data shown in \textbf{Fig.~4a}, the
proportionality $\Omega^2 = \beta P_0$ between excitation power
$P_0$ and effective Rabi frequency has been substituted in the
above expression. By consideration of the independently verified
values of $T_1$ (radiative life time) and coherence $T_2 = 2 \hbar
/ \Gamma_0$ (with $\Gamma_0$ as the zero-power limit of emission
linewidth), high conformity is achieved with the experimental mode
emission saturation behaviour, which therefore reflects the
characteristics of the coupled, resonantly pumped QD. We like to
emphasize that the exact value of $T_1$ strictly depends on the
actual Purcell enhancement, i.e. the $\Delta E$ detuning-dependent
emitter-mode coupling. In temperature-dependent tuning
experiments, the value of $T_2$ is influenced by the effect of
pure dephasing by phonon coupling which increases the emission
linewidth with increasing temperature $T$.

%Second-Order Correlations: Non-resonant vs. resonant excitation

Investigations on the photon emission statistics have been
performed under the conditions of either (a) \textit{off-resonant}
excitation into the first excited (p-shell) state of single QDs
(see \textbf{Fig.~1c - e} as well as (b) \textit{purely resonant}
excitation into the QD s-shell (\textbf{Fig.~4b}). In either case,
the raw correlation data is normalized to the expectation value of
correlation events for a Poisson-distributed source, i.e.
$N_{Poisson}^{cw} = N_{start} \cdot N_{stop} \cdot \Delta t_{int}
\cdot \Delta t_{MCA}$ \cite{Brouri.Beveratos:2000}. In this
expression, $N_{start,stop}$ correspond to the detector count
rates, $\Delta t_{int}$ represents the total time of integration,
and $\Delta t_{MCA}$ is the temporal bin width of a single channel
of the multi-channel analyzer used for accumulation.

(a) As is demonstrated by the dashed lines in \textbf{Figs.~1c -
e} for the case of \textit{off-resonant excitation}, the
Poisson-normalized correlation data traces can be described by a
simple theoretical expression \cite{Becher.Kiraz:2001} of the form
\begin{equation}
g^{(2)}(\tau) = 1 - \rho^2 \exp{\left ( -|\tau| / t_m \right )}
\quad .
\end{equation}
In this notation, $\rho = S/(S+B)$ represents the
signal-to-background ratio of the detected $\mu$-PL signal, $t_m$
is the (excitation power-dependent) anti-bunching time constant,
and $\tau = t_{stop} - t_{start}$ is the measured photon-photon
delay, respectively. In order to additionally account for the
limited temporal resolution if the detection system, the above
expression has to be convoluted by the instrumental response
function, which was assumed Gaussian-shaped with $2 \sigma \approx
\Delta t_{IRF} = 400$~ps in our case. As is discussed in the text,
the convoluted function (bold traces) reveals good quantitative
agreement with the experimental $g^{(2)}(\tau)$ data, allowing for
a correct interpretation of the quality of single-photon
generation.

(b) Under the conditions of \textit{strictly resonant QD s-shell
excitation} \cite{Wrigge.Sandoghdar:2008} and in the limits of
\textit{weak optical pumping}, the Poisson-normalized correlation
is theoretically expressed by
\begin{equation}\label{eqn:resg2}
g^{(2)}_{res}(\tau) = 1 - \left( 1 - \frac{T_1}{T_2} \right)^{-1}
\cdot \exp \left( -\frac{|\tau|}{T_2} \right) - \left( 1 -
\frac{T_2}{T_1} \right)^{-1} \cdot \exp \left( -\frac{|\tau|}{T_1}
\right) \quad .
\end{equation}
In this weak power regime, no Rabi oscillations are expected. In
the above expression, $T_1$ and $T_2$ represent the radiative
lifetime and the emission coherence time of the transition,
respectively. A corresponding fit (dashed line) has been applied
to our experimental data in \textbf{Fig.~4b}, which reflects
single QD resonance emission coupled to a pillar mode. In order to
analyze the influence of limited time resolution, also here a
convolution with the instrumental response $\Delta t_{IRF}$ (bold
line) is applied in comparison with the idealized fit of
\textbf{Eq.~\ref{eqn:resg2}}. As is discussed in the main text, we
could verify the ideal conditions of background-free single-photon
emission from the measured mode signal, which fully reflects the
expected emission statistics of the coupled QD.

\end{document}